\documentclass[twocolumn,superscriptaddress,longbibliography,
	aps,pra,preprintnumbers]{revtex4-2}

\usepackage[normalem]{ulem}
\usepackage{graphicx}
\usepackage{bm}
\usepackage{color}
\usepackage{epstopdf}
\usepackage{amsmath}
\usepackage{amssymb}
\usepackage{epstopdf}
\usepackage{lipsum}

\usepackage{gensymb}

\usepackage{multirow}

\usepackage{scrextend}

\usepackage[urlcolor=blue,colorlinks=true,citecolor=blue,linkcolor=blue,pdfstartview={FitH},bookmarks=false]{hyperref}

\usepackage[dvipsnames]{xcolor}
\definecolor{mygreen}{rgb}{0.0, 0.6, 0.0}
\definecolor{pjorange}{rgb}{0.8, 0.3, 0.0}
\definecolor{jlblue}{rgb}{0.2, 0.5, 0.7}

\graphicspath{{fig/}{./fig/}{.}}

\begin{document}

\title{
Lattice dynamics of altermagnetic ruthenium oxide RuO$_{2}$
}

\author{Surajit Basak}
\email[e-mail: ]{surajit.basak@ifj.edu.pl}
\affiliation{\mbox{Institute of Nuclear Physics, Polish Academy of Sciences, W. E. Radzikowskiego 152, PL-31342 Krak\'{o}w, Poland}}

\author{Andrzej~Ptok}
\email[e-mail: ]{aptok@mmj.pl}
\affiliation{\mbox{Institute of Nuclear Physics, Polish Academy of Sciences, W. E. Radzikowskiego 152, PL-31342 Krak\'{o}w, Poland}}

\date{\today}

\begin{abstract}
Altermagnetic ruthenium oxide RuO$_{2}$ crystallizes with P4$_{2}$/mnm symmetry.
Here we discuss the lattice dynamics of this structure. 
We show and discuss the phonon dispersion and density of states. 
The phonon dispersion curves contain several Dirac nodal lines and highly degenerate Dirac points.
We present the characteristic frequencies and their irreducible representations at the $\Gamma$ point.
Theoretically obtained frequencies of the Raman active modes nicely reproduce the ones reported experimentally.
\end{abstract}

\maketitle

\section{Introduction}
\label{sec.intro}

The altermagnetic is a new elementary phases of magnetically ordered systems~\cite{smejkal.sinova.22,smejkal.sinova.22b}.
This phase is characterized by the combined features of ferromagnetic (FM) and antiferromagnetic (AFM), leading to novel effects.
Altermagnetic phase breaks time reversal symmetry similar to the FM phase, and possess compensated magnetization like the AFM phase.
Unlike ferromagnets, however, the altermagnetic spin splitting in the nonrelativistic bands is accompanied by a symmetry-protected zero net magnetization~\cite{gopalan.litvin.11}.
The spin split part of the band structure is accompanied by spin degeneracies along certain surfaces in the Brillouin zone.

\begin{figure}[!b]
\centering
\includegraphics[width=\columnwidth]{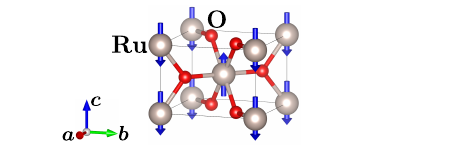}
\caption{
The crystal structure of altermagnetic ruthenium oxide RuO$_{2}$.
\label{fig.schemat}
}
\end{figure}

Presently, several compounds realizing the altermagnetic order are known (for more details see Ref.~\cite{guo.liu.23}).
In our paper we discuss the dynamical properties of RuO$_{2}$~\cite{smejkal.gonzalezhernandez.20}, crystallized with the rutile structure (see Fig.~\ref{fig.schemat}).
The electronic band structure exhibits spin splitting in range of $0.5$~eV~\cite{zhan.li.23}.
Such spin band splitting should be reflected in the spin-polarized angle resolved photoelectron spectroscopy (ARPES) measurement~\cite{ptok.23}.
The time reversal symmetry breaking~\cite{fedchenko.minar.23} and the anomalous Hall effect~\cite{feng.zhou.22} were observed experimentally.
Both phenomena are related to the topological properties, associated with non-zero Berry phase~\cite{smejkal.macdonald.22}.
Finally, thin film of RuO$_{2}$ exhibits superconducting properties~\cite{ruf.paik.21}.

Altermagnetic RuO$_{2}$ posses also several properties, interesting from application point of view, e.g. high activation barrier~\cite{torun.fang.13}.
This compound is a prime catalyst for the oxygen evolution reaction in water splitting~\cite{liang.bieberlehutter.22}.
In the context of thermoelectric properties~\cite{music.kremer.15}, the dynamical properties of RuO$_{2}$ can be interesting.
Here, we present and discuss the dynamical properties of the altermagnetic RuO$_{2}$.

The paper is organized as follows.
The computational details are presented in Sec~\ref{sec.comp}.
Next, in Sec.~\ref{sec.res}, we present and discuss our numerical results.
Finally, a summary is included in Sec.~\ref{sec.sum}.

%%%%%%%%%%%%%%%%%%%%%%%%%%%%%%%%%%%%%%%%%%
%%%%%%%%%%%%%%%%%%%%%%%%%%%%%%%%%%%%%%%%%%
%%%%%%%%%%%%%%%%%%%%%%%%%%%%%%%%%%%%%%%%%%

\section{Computational details}
\label{sec.comp}

The first-principles density functional theory (DFT) calculations were performed using the projector augmented-wave (PAW) potentials~\cite{blochl.94} implemented in the Vienna Ab initio Simulation Package ({\sc Vasp}) code~\cite{kresse.hafner.94,kresse.furthmuller.96,kresse.joubert.99}.
For the exchange-correlation energy, the generalized gradient approximation (GGA) in the Perdew, Burke, and Ernzerhof (PBE) parametrization was used~\cite{pardew.burke.96}.
Similar to the previous study~\cite{ptok.23}, we apply the correlation effects on Ru $d$ orbitals within DFT+U approach, introduced by Dudarev {\it et al.}~\cite{dudarev.botton.98}.
The energy cutoff for the plane-wave expansion was set to $600$~eV.
Optimization of the structural parameters (in the presence of the spin--orbit coupling) was performed using $10 \times 10 \times 15$ ${\bm k}$--point grid using the Monkhorst--Pack scheme~\cite{monkhorst.pack.76}.
As a convergence condition of the optimization loop, we took the energy change below $10^{-6}$~eV and $10^{-8}$~eV for ionic and electronic degrees of freedom respectively.

The dynamical properties were calculated using the direct {\it Parlinski-Li-Kawazoe} method~\cite{parlinski.li.97} implemented in {\sc Phonopy}~\cite{togo.tanaka.15}.
The interatomic force constants were found from the force acting on the atoms displaced from the equilibrium position.
In these calculations, we used the supercell containing $2 \times 2 \times 3$ primitive unit cells, and reduced $4 \times 4 \times 4$ ${\bm k}$-grid.

%%%%%%%%%%%%%%%%%%%%%%%%%%%%%%%%%%%%%%%%%%
%%%%%%%%%%%%%%%%%%%%%%%%%%%%%%%%%%%%%%%%%%
%%%%%%%%%%%%%%%%%%%%%%%%%%%%%%%%%%%%%%%%%%

\section{Results and discussion}
\label{sec.res}

\subsection{Crystal structure}

RuO$_{2}$ crystallizes with the rutile structure (P4$_2$/mnm, space group No.~136) presented in Fig.~\ref{fig.schemat}.
It has one Ru atom sitting at each corner of a unit cell, as well as one Ru atom at the center.
Each Ru atom is surrounded by six O atoms that form a distorted octahedron.
In this case, the Ru and O atoms are located at Wyckoff positions $2b$ (0,0,0) and $4g$ ($x_\text{O}$,$y_\text{O}$,0), respectively.
Here, $x_\text{O}$ and $y_\text{O}$ are two free parameters describing position of O atoms in the crystal structure.

Theoretically, the obtained crystal parameters weakly depend on the assumed $U$~\cite{ptok.23}, and in practice $U$ affects only on magnetic moment of Ru atoms.
Similar to the previous study~\cite{ahn.hariki.19}, we assume $U = 2$~eV for Ru $d$ orbitals. 
In this case the magnetic moment of Ru is equal $1.152$~$\mu_{B}$.
After optimization we find $a = 4.533$~\AA, $c = 3.11$~\AA, while the free parameters are obtained as $x_\text{O} = 0.8037$ and $y_\text{O} = 0.1963$. 
The lattice constants are close to the experimentally observed ones, i.e., $a = 4.49$~\AA, and $c = 3.11$~\AA~\cite{zhu.strempfer.19}.

\begin{figure}[!b]
\centering
\includegraphics[width=\columnwidth]{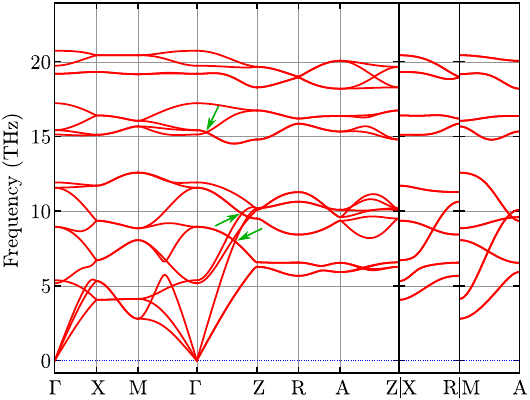}
\caption{
The phonon dispersion curves of altermagnetic ruthenium oxide RuO$_{2}$.
Green arrows marks the fourfold Dirac points along $\Gamma$--Z.
\label{fig.ph_band}
}
\end{figure}

\begin{figure}[!t]
\centering
\includegraphics[width=\columnwidth]{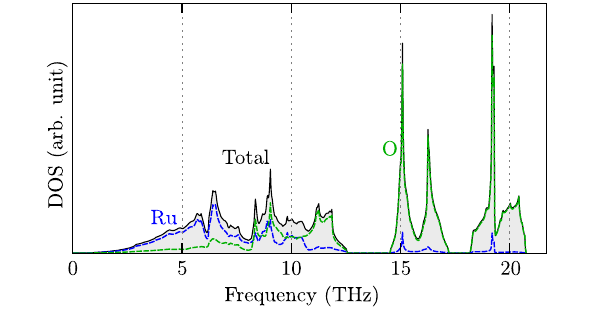}
\caption{
The phonon density of states of altermagnetic ruthenium oxide RuO$_{2}$.
\label{fig.ph_dos}
}
\end{figure}

\subsection{Phonon dispersion and density of states}

The phonon dispersion curves are presented in Fig.~\ref{fig.ph_band}, and are similar to the one reported earlier~\cite{uchida.nonoto.20}.
Corresponding phonon density of states (DOS) is presented in Fig.~\ref{fig.ph_dos}.
The vibrations in low frequency range (smaller than $7.5$~THz) are mostly related to the Ru atoms.
Similarly, higher frequency vibrations (above $14$~THz) are mostly associated with vibrations of the lighter atoms, i.e. oxygen.
Interestingly, vibration modes of oxygen atoms at higher energies are separated from the rest of modes by a gap.
This frequency gap is well visible in the phonon band structure (Fig.~\ref{fig.ph_band}).

In the phonon band structure we can find several interesting features, well known from the electronic band structure of RuO$_{2}$~\cite{jovic.koch.18} or non-magnetic IrO$_{2}$ with the same structure~\cite{xu.jiang.19}.
The doubly degenerate Dirac nodal lines (DNL) along M--X path originate from the mirror symmetry along the ($110$) and ($\bar{1}10$) planes.
Similarly mirror symmetry along ($001$) plane, lead to double degeneracy of bands for R--A path.
Additionally, nonsymmorphic symmetry including a fourfold  screw  rotation  around  the $z$ axis $C_{4z}$ and a glide mirror reflection along the ($100$) and ($010$) planes, also allows the realization of doubly degenerate DNL along $\Gamma$--Z and M--A.

Degeneracy of bands along $\Gamma$--Z is the same as their degeneracy at $\Gamma$ point (see also Sec.~\ref{sec.active}).
Along M--A all the phonon bands are doubly degenerate.
Additionally, along $\Gamma$--Z, highly degenerate Dirac points can be realized, coming from the crossing of single- or doubly degenerate bands.
This leads to triple- and fourfold degenerate Dirac points (few fourfold degenerate Dirac points are marked by green arrows in Fig.~\ref{fig.ph_band}).

\subsection{IR and Raman active modes}
\label{sec.active}

The phonon modes at the $\Gamma$ point can be decomposed into irreducible representations of the space group $P4_{2}/mnm$ as follows: 
\begin{eqnarray}
\nonumber \Gamma_\text{acoustic} &=& A_\text{2u} + E_\text{u} \\
\Gamma_\text{optic} &=& A_\text{1g} + A_\text{2g} + A_\text{2u} \\
\nonumber &+& B_\text{1g} + 2 B_\text{1u} + B_\text{2g} + 3 E_\text{u} + E_\text{g}
\end{eqnarray}
In total, there are 18 vibrational modes, eight nodegenerate ($A_\text{1g}$, $A_\text{2g}$, $2 A_\text{2u}$, $B_\text{1g}$, $B_\text{2g}$, and $2 B_\text{1u}$) and ten doubly degenerate ($E_\text{g}$ and $4 E_\text{u}$).
Among these, $A_\text{2u}$ and $E_\text{u}$ are infrared (IR) active, while $A_\text{1g}$, $B_\text{1g}$, $B_\text{2g}$, and $E_\text{g}$ are Raman active.
Raman active modes are related only to the oxygen atoms displacement.
Contrary to this, IR active modes are related with displacement of both Ru and O atoms.

\begin{table}[!t]
\caption{
\label{tab.ramanselect}
Selection rules for Raman-active modes.
}
\begin{ruledtabular}
\begin{tabular}{lcccc}
Configuration & $A_\text{1g}$ & $B_\text{1g}$ & $B_\text{2g}$ & $E_\text{g}$ \\
\hline 
$e_{x}$ in $e_{x}$ out (linear $\parallel$) & $|a|^{2}$ & $|c|^{2}$ & 0 & 0 \\
$e_{x}$ in $e_{y}$ out (linear $\perp$) & $0$ & 0 & $|d|^{2}$ & 0 \\
$e_{x}$ in $e_{z}$ out (linear $\perp$) & $0$ & 0 & 0 & $|e|^{2}$ \\
\end{tabular}
\end{ruledtabular}
\end{table}

\paragraph*{Selection rules for Raman-active modes. ---}
The non-resonant Raman scattering intensity depends in general on the directions of the incident and scattered light relative to the principal axes of the crystal. 
It is expressed by Raman tensor $R$, relevant for a given crystal symmetry, as~\cite{loudon.01}:
\begin{eqnarray}
\label{eq.intens} I \propto | e_{i} \cdot R \cdot e_{s} |^{2} ,
\end{eqnarray}
where $e_{i}$ and $e_{s}$ are the polarization vectors of the incident and scattered light, respectively.
According to group theory, the Raman tensor for the $P4_{2}/mnm$ space group takes the following forms for the Raman active modes:
\begin{eqnarray}
\label{eq.raman} &R \left( A_\text{1g} \right) = & \left( \begin{array}{ccc}
a & 0 & 0 \\ 
0 & a & 0 \\ 
0 & 0 & b
\end{array} \right) ; \\
\nonumber &R \left( B_\text{1g} \right) = &
\left( \begin{array}{ccc}
c & 0 & 0 \\ 
0 & -c & 0 \\ 
0 & 0 & 0
\end{array} \right) ; \; R \left( B_\text{2g} \right) =
\left( \begin{array}{ccc}
0 & d & 0 \\ 
d & 0 & 0 \\ 
0 & 0 & 0
\end{array} \right) ; \\
\nonumber &R \left( E_\text{g}^\text{I} \right) =& \left( \begin{array}{ccc}
0 & 0 & 0 \\ 
0 & 0 & e \\ 
0 & e & 0
\end{array} \right); \;\;\; R \left( E_\text{g}^\text{II} \right) =
\left( \begin{array}{ccc}
0 & 0 & -e \\ 
0 & 0 & 0 \\ 
-e & 0 & 0
\end{array} \right) . 
\end{eqnarray}
Using Eq.~(\ref{eq.intens}) and the Raman tensors~(\ref{eq.raman}), we can determine the selection rules and Raman intensities for various scattering geometries.
Tab.~\ref{tab.ramanselect} summarized the Raman response in the backscattering geometry for four polarization configurations.
As we can see, it is possible to distinguish the Raman active modes using the different backscattering configurations.
In this context, the observed Raman modes possess different insensitivity, as shown in Table.~\ref{tab.ramanselect}.
Indeed, this technique was successfully used experimentally to recognize the Raman active modes~\cite{chen.korotcov.07,kim.baik.10}.

Theoretically obtained characteristic frequencies of the modes at $\Gamma$ point and their irreducible representations for RuO$_{2}$ are collected in Table~\ref{tab.irr}.
The frequencies of the Raman active modes at higher frequencies are underestimated with respect to the experimentally reported ones~\cite{chen.korotcov.07}.
Moreover, the ``bulk'' Raman active modes have the same frequencies like the ones reported for RuO$_{2}$ nanowires~\cite{kim.baik.10}.
Finally, we should note that the observed frequencies can strongly depend on the experimental setup, e.g. ``strain''~\cite{chen.korotcov.07,uchida.nonoto.20}.

\section{Summary}
\label{sec.sum}

In this paper we discussed the lattice dynamic of altermagnetic ruthenium oxide RuO$_{2}$.
We show that the bulk system is stable with $P4_{2}/mnm$ symmetry.
The phonon dispersion curves are well defined, i.e. all phonon branches possess real frequencies.
The vibrations at low frequency range are related to the Ru modes, while the branches at highly frequency range are mostly associated with vibration of O atoms.
We discussed also the characteristic frequencies of the modes at $\Gamma$ point and selective rules of Raman active modes.
The Raman active mode are distinguished within backscattering geometry measurements.
Theoretically obtained frequencies of the Raman active modes nicely reproduce the experimentally reported data.

\begin{table}[!t]
\caption{
\label{tab.irr}
Symmetries of irreducible representations and their characteristic frequencies at $\Gamma$ point for RuO$_{2}$.
The experimental frequencies of the Raman active modes correspond to the single crystal measurements presented in Ref.~\cite{chen.korotcov.07}.
}
\begin{ruledtabular}
\begin{tabular}{cccc}
Symm. & Freq. (THz) & Freq. (cm$^{-1}$) & Exp.\footnote{Frequencies only for Raman active modes} (cm$^{-1}$) \\
\hline
$B_\text{1g}$ &  $5.180$ & $172.79$ & N/A \\
$B_\text{1u}$ &  $5.375$ & $179.29$ & \\
$E_\text{u}$  &  $8.951$ & $298.57$ & \\
$E_\text{u}$  & $11.569$ & $385.90$ & \\
$A_\text{2g}$ & $11.924$ & $397.74$ & \\
$A_\text{2u}$ & $15.131$ & $504.72$ & \\
$E_\text{g}$  & $15.420$ & $514.36$ & $526$ \\
$B_\text{1u}$ & $17.225$ & $574.57$ & \\
$E_\text{u}$  & $19.193$ & $640.22$ & \\
$A_\text{1g}$ & $19.735$ & $658.29$ & $644$ \\
$B_\text{2g}$ & $20.736$ & $691.68$ & $714$  
\end{tabular}
\end{ruledtabular}
\end{table}

\begin{acknowledgments}
%The author thanks \AP{xxx} for helpful discussions and critical reading of the manuscript. 
Some figures in this work were rendered using {\sc Vesta}~\cite{momma.izumi.11}.
We kindly acknowledge support by National Science Centre (NCN, Poland) under Project No.~2021/43/B/ST3/02166.
\end{acknowledgments}

%%%%%%%%%%%%%%%%%%%%%%%%%%%%%%%%%%%%%%%%%%
%%%%%%%%%%%%%%%%%%%%%%%%%%%%%%%%%%%%%%%%%%
%%%%%%%%%%%%%%%%%%%%%%%%%%%%%%%%%%%%%%%%%%
%%%%%%%%%%%%%%%%%%%%%%%%%%%%%%%%%%%%%%%%%%

%\nocite{*}
\bibliography{biblio.bib}

\end{document}